\documentclass{mem}
\usepackage{natbib}\usepackage{txfonts}\usepackage{balance}
\usepackage{graphicx}
\usepackage[a4paper,breaklinks,dvipdfm]{hyperref}
\idline{83}{1}
\begin{document}
\def\teff{$T\rm_{eff }$}
\def\kms{$\mathrm {km s}^{-1}$}

\title{
Progress in Understanding\\ the Diffuse UV Cosmic Background
}

   \subtitle{}

\author{
Richard Conn Henry 
          }

  \offprints{R. C. Henry}

\institute{
Henry A. Rowland Department of Physics \& Astronomy,
The Johns Hopkins University,
Baltimore, Maryland, 21218-2686 USA\\
\email{henry@jhu.edu}
}

\authorrunning{Henry }

\titlerunning{Diffuse UV Background}

\abstract{
I report on progress in my ongoing work with Professor Jayant Murthy 
concerning the origin and nature of the diffuse ultraviolet background radiation over 
the sky.  We have obtained and are reducing a vast trove of Voyager 
ultraviolet spectrometer observations of the diffuse background {\it shortward} 
of L$\alpha$, including for the first time measurements made
from the outermost regions of the solar system, where noise from solar-system scattered (and then
 grating-scattered)
 solar L$\alpha$ is lowest.  Also, we have obtained and are investigating
the complete set of GALEX observations of the diffuse ultraviolet background
{\it longward} of L$\alpha$.  Preliminary investigation appears to confirm that longward of L$\alpha$ there
exists a component of the diffuse ultraviolet background that is not dust-scattered starlight.
\keywords{techniques:  ultraviolet --- cosmology: diffuse background }
}
\maketitle{}

\section{Introduction}

\begin{figure*}[t!]  
\resizebox{\hsize}{!}{\includegraphics[clip=true]{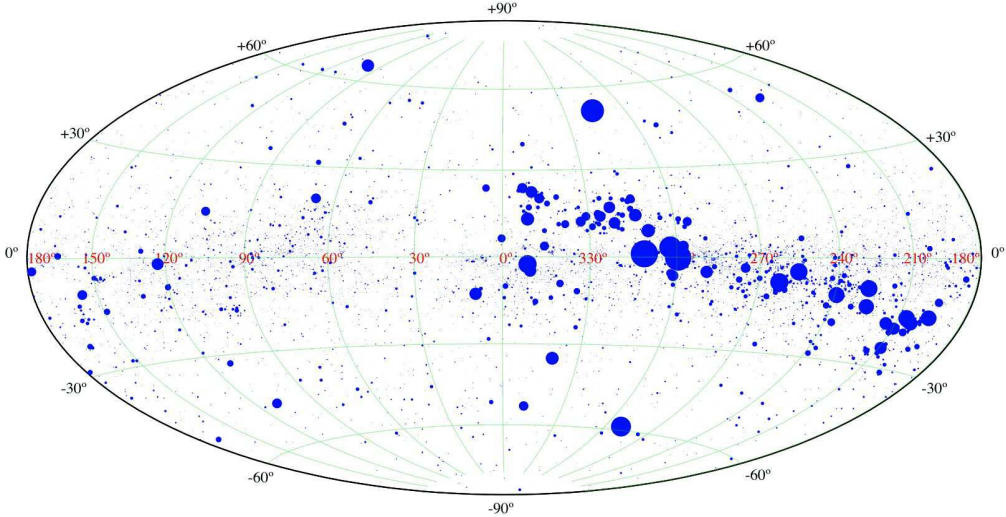}}
\caption{\footnotesize
The asymmetry in the $(\ell, b)$ distribution of these, the UV stars, should permit easy separation of
 dust-scattered starlight from any additional galactic-symmetric diffuse radiation source that may exist.
 Note Spica ($\ell$=316.11, $b$=+50.85), $\alpha$~Eri (290.84, -58.79), and $\eta$ U Ma ($\ell$=100.69, $b$=+65.32
 , $d$=30.9$\pm$0.7 pc).
}
\label{fig-1}
\end{figure*}
\begin{figure}[t!] 
\resizebox{\hsize}{!}{\includegraphics[clip=true]{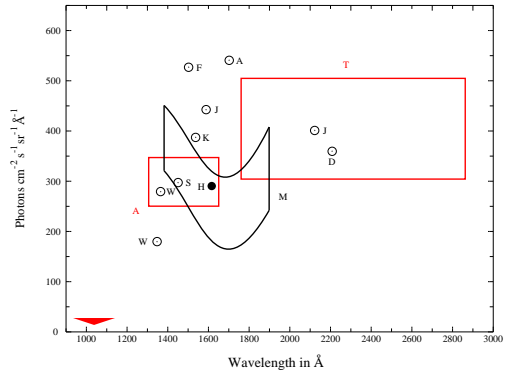}}
\caption{
\footnotesize
This figure, adapted from Henry \& Murthy (1995), identifies the mystery that it is
our goal is resolve:  what is the origin of the abrupt drop in the high-galactic-latitude diffuse ultraviolet background
that occurs shortward of approximately 1200~\AA?  The drop was discovered by Holberg (1986).
}
\label{fig-2}
\end{figure}

I have previously reported, at these Frascati Vulcano workshops (1997,
 1999, 2001, 2003, 2005, 2009), on my progress
in understanding the diffuse UV cosmic background.

The celestial map in Fig.~1 shows the unextincted {\it direct} stellar
flux (proportional to the area of the symbol) at $1500\, \AA$ from each of the 31215 TD1
stars that are expected (Bowyer 1991) to source the bulk of the diffuse UV background
radiation, through the scattering of their light by interstellar dust---but I am searching 
for a component that may be present {\it in addition} to that predictable source (Henry 1991).

That such an additional component may indeed exist was first suggested by the spectral character
of the observed diffuse UV background  as shown in Fig.~2, which displays observations made over many years using
many instruments.  We see that longward of approximately L$\alpha$ there
is a substantial background at high galactic latitudes, but that shortward of that wavelength,
as revealed by our Voyager observations (Murthy et al. 1999), there is no
detectable general diffuse UV background.  Barring an unphysical
abrupt change in interstellar grain albedo with wavelength, the radiation longward of  L$\alpha$ has been
revealed to be, at high galactic latitudes, largely {\it not} dust-scattered starlight.

I begin a further test of this idea, by comparing the observed GALEX FUV background (Fig.~3) with 
the first prediction of my new (and extremely simple) model (Fig. 4) of dust-scattered starlight.  
My model is generated from the 
stars of Fig.~1 (the nearer ones placed at Hipparcos distances) traversing dust with a scale height of (in this first trial case)
 100 pc above and below the 
galactic plane, and with assumed grain albedo of 0.1 and grain Henyey-Greenstein scattering parameter
$g$ = 0.58. 

In my models, for a given location on the sky, and distance, the light from each star is extincted,
on its way to that location---with the scattered part being simply directly forward scattered, the 
absorbed portion of course disappearing.  On arrival at the observed line of sight, absorption again occurs,
but this time the scattered light is scattered properly, with a Henyey-Greenstein scattering
function, and some fraction goes in the direction of the GALEX camera. Finally, that scattered light is 
again absorbed/forward-scattered on its  way to the GALEX camera.

\section{Observed vs. Predicted}

The maps of Figs. 3 and 4, which are filled in only for the portions of the sky that were observed with GALEX, should be carefully compared:   Fig.~3, the {\it observed} GALEX brightnesses (adapted from Murthy, Henry, \& Sujatha 2010), clearly shows the dust-scattered starlight that is expected from the stars of Fig.~1 (and which I have
 {\it predicted} in Fig.~4).  But, it also shows clear evidence for what seems to be an additional component
of diffuse radiation: one that is symmetric with the galactic plane.

\section{Individual Stars}

Murthy \& Henry (2011) have used some of the data that are shown in Fig. 3 to obtain a robust measurement, at 1530 $\AA$, of the Henyey-Greenstein scattering parameter  $g$=0.58$\pm$0.12.  They accomplished this by studying the 
scattered light from individual  stars such as Spica and  $\alpha$~Eri (Fig. 1).

\begin{figure*}[t!]   
\resizebox{\hsize}{!}{\includegraphics[clip=true]{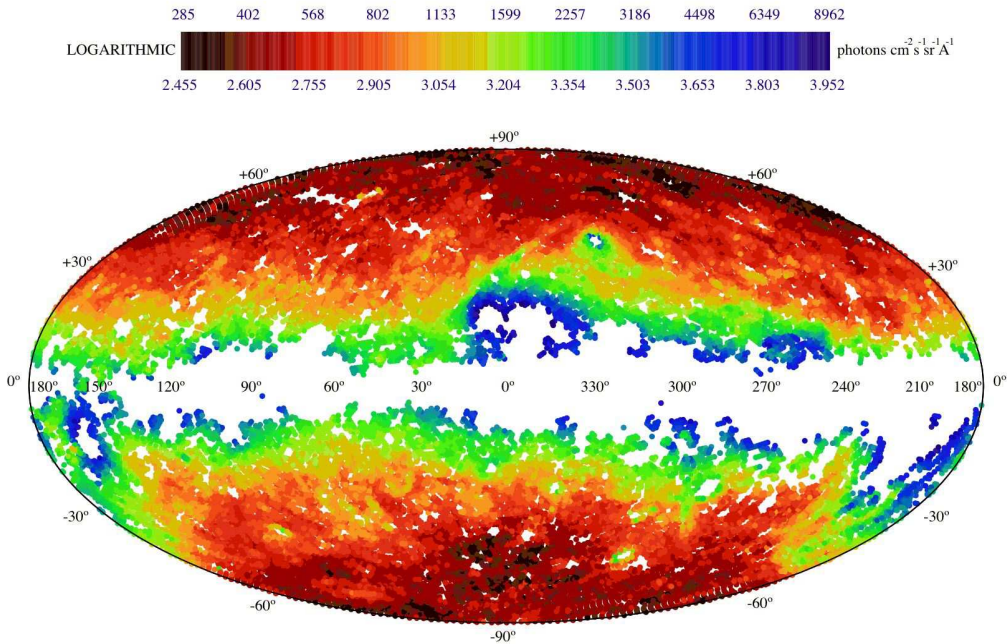}}
\caption{\footnotesize
Observed GALEX diffuse background radiation at $1530\, \AA$.  We detect diffuse UV 
light from Spica ($\ell$=316.11, $b$=+50.85).  There is significant diffuse UV at low latitudes in  the
range 15$^{\circ}$ to 45$^{\circ}$ where there are no bright source stars (see Fig. 1).  There is no diffuse UV detected from
$\eta$ U Ma ($\ell$=100.69, $b$=+65.32, d=30.9$\pm$0.7 pc), but there is plenty of radiation at high galactic latitudes that 
is present independent of galactic longitude.  (Regions that are blank were not observed with GALEX.)
}
\label{fig-3}
\end{figure*}
\begin{figure*}[t!]    
\resizebox{\hsize}{!}{\includegraphics[clip=true]{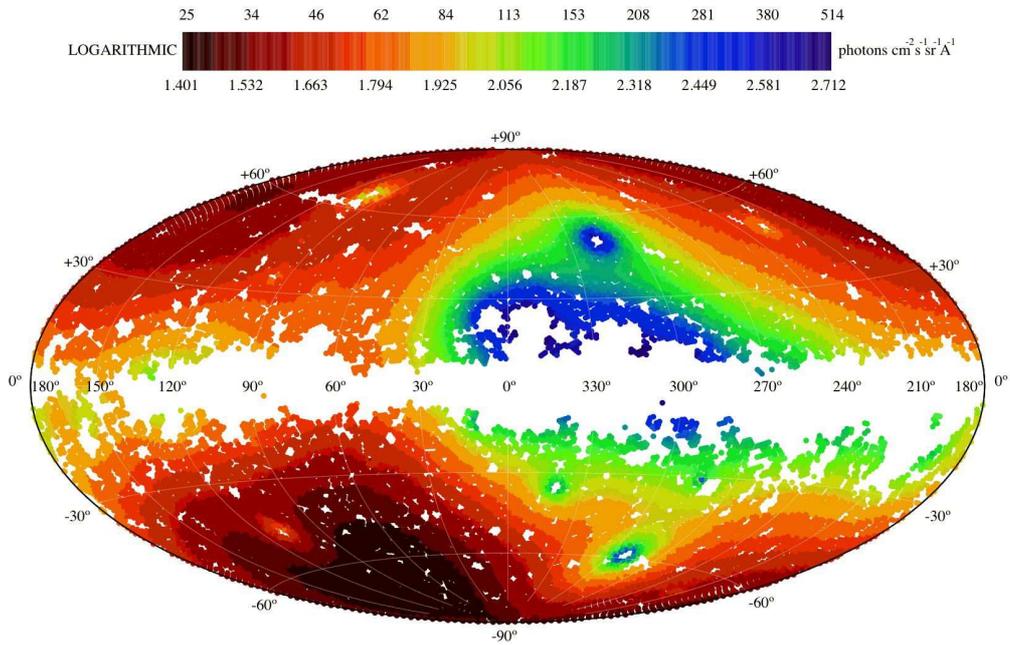}}
\caption{\footnotesize
The prediction of my simple model, with $g$ = 0.58, of the diffuse (only) sky brightness at $1530\, \AA$.
}
\label{fig-4}
\end{figure*}
\begin{figure*}[t!]   
\resizebox{\hsize}{!}{\includegraphics[clip=true]{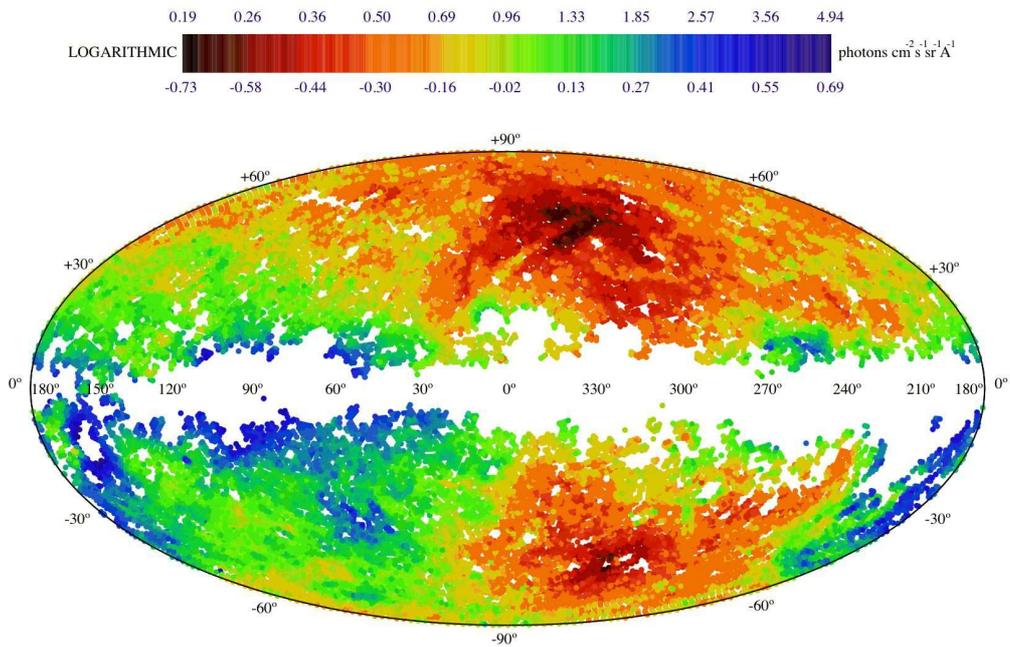}}
\caption{\footnotesize
(FUV observed, Fig. 3)/(FUV model, Fig. 4), normalized at peak brightnesses.  Much greater asymmetry
with galactic longitude is seen than appears in the observations of Fig. 3.
}
\label{fig-5}
\end{figure*}
\begin{figure}[t!]   
\resizebox{\hsize}{!}{\includegraphics[clip=true]{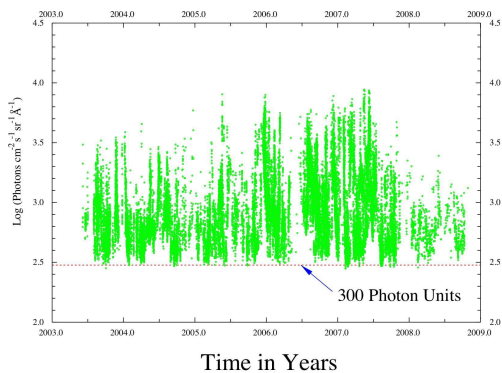}}
\caption{
\footnotesize
The UV brightness of 31902 GALEX diffuse backgrounds versus time: no solar-cycle effect.
}
\label{fig-6}
\end{figure}
\begin{figure*}[t!]    
\resizebox{\hsize}{!}{\includegraphics[clip=true]{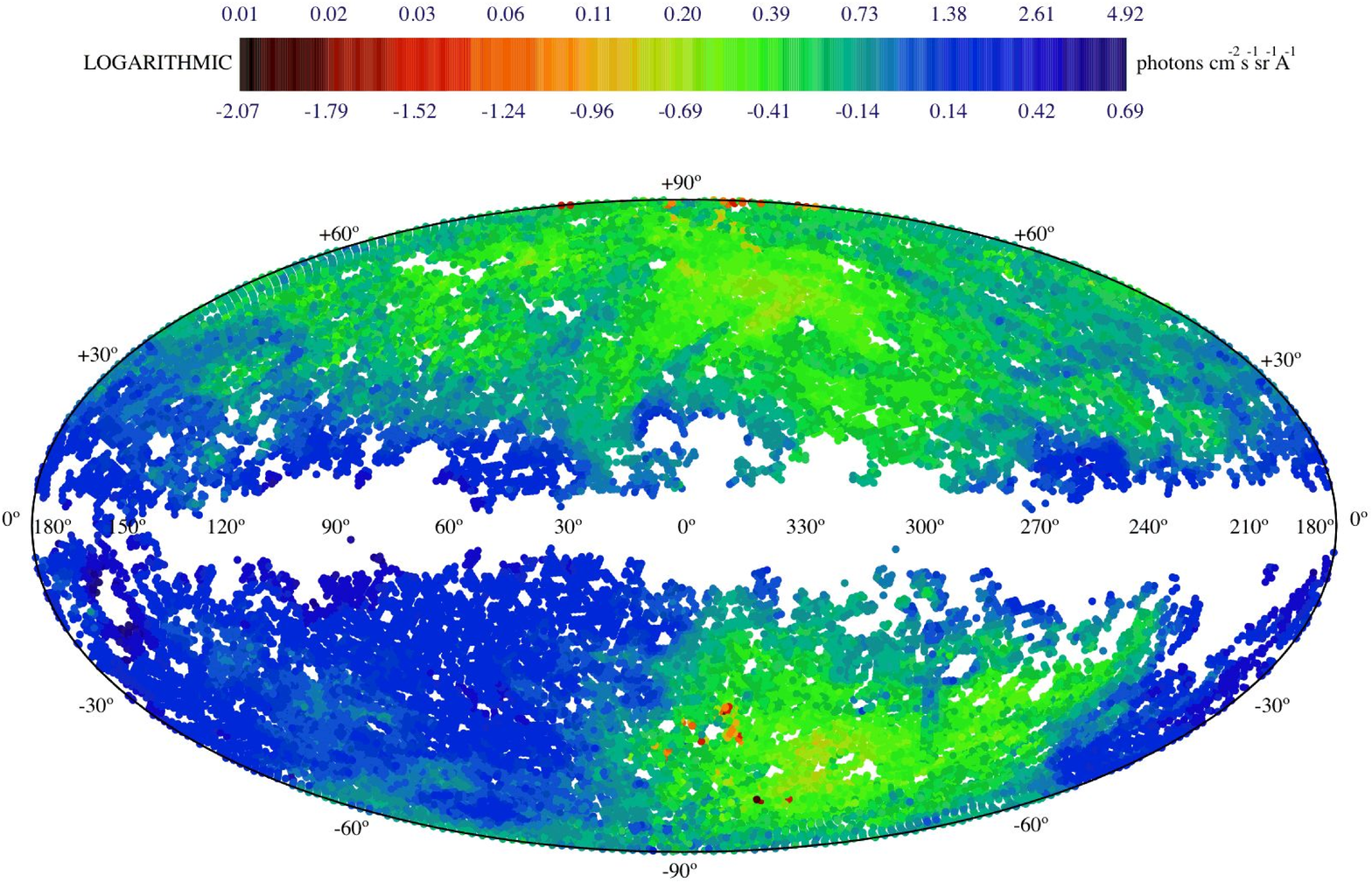}}
\caption{\footnotesize
(FUV observed, but now corrected for airglow)/(FUV model), normalized at peak brightnesses.
Correction artifacts (black, red spots) distort the intensity scale, but qualitative accord with Fig. 6 is good.
}
\label{fig-7}
\end{figure*}

\section{Discussion}

My model prediction (Fig.~4) shows a range from 25 to 514 photons cm$^{-2}$ s$^{-1}$ sr$^{-1}$ \AA$^{-1}$,
which is rather drastically different from the observed (Fig.~3) range (285 to 8962 units)!

Of course I have not yet started varying the parameters (scale height, grain albedo, 
scattering parameter, amount of dust).  A run of one model takes about 24 hours on my Macintosh computer.
I expect that I will be able to do better, as I progress, in matching prediction to
observation.  But, given the extreme simplicity of the model,
it is extraordinary how
well it does, even on this very first run, in matching the observations, in terms of distribution on the 
sky---see Fig.~5, which shows the data of Figs.~3 and 4 in normalized ratio:
 (observed/model)$\times$(514/8962).  That is, I force agreement at the brightest 
locations, which are surely dominated by dust-scattered starlight.
The result is extremely satisfying, and suggests that this approach might, with further work, even more convincingly tease out
our suspected second component.  
Note the scale in Figure 5:  the range of the deviations is
remarkably small:  there are, on the map, no spots that are more than a factor five brighter than they ``should be,"
and no spots that are less than 1/5 as bright as they ``should be," {\it were} dust-scattered starlight 
the {\it sole} source of the observed radiation.  All this is relative, of course, but it is over the entire set of observations!
For a first try, I regard this result as being extremely satisfactory:  we seem to
 have a robust handle on the dust-scattered-starlight
component of what GALEX observes!

Even at this preliminary stage, I think that important conclusions are possible.  The observations (Fig. 3) show,
at galactic latitudes above $\pm 30^{\circ}$, no dependence on galactic longitude, despite the drastic dependence  
on galactic longitude of
the putative source (stars), as clearly appears in Fig.~1. Fig.~5 shows a {\it strong} longitude dependence in the
predicted brightness of dust-scattered starlight!  In particular, our adoption
 of Henyey-Greenstein $g$=0.58 (Murthy \& Henry 2011) {\it under}predicts
what is seen.  This means that the (non-physical) Henyey-Greenstein (1941) function is incorrect, which is no surprise.  But
it clearly seems impossible that the diffuse background above latitude $\pm 30^{\circ}$ can originate in starlight scattered
from interstellar dust. That is the main conclusion of this paper.

\section{Corrections to the Data}

What appears in Fig. 3 is the raw observed diffuse background, which is known to be contaminated to
some degree:  as the spacecraft
begins each observation, on leaving the sunlit side of the Earth, the observed celestial brightness slowly decreases, with
time, to a minimum, and then rises again as the GALEX spacecraft approaches the end of the night time
portion of its orbit---the time-variable portion of the signal is of course contamination of some kind.  In Fig. 6, I demonstrate that
 whatever the source of that contamination, it is independent of the solar cycle.The contamination is not large,
but one must be concerned about its effects.

Murthy, Henry, and Sujatha (2010) give {\it two} values for the diffuse UV background for each GALEX observation:  the
raw observation (as I used in all previous figures), and corrected as best we could for time-variable contamination.

How important is such contamination?  The reader can tell to some degree by examining Fig. 7, which is identical to
Fig. 5 except that now I have used the corrected (instead of the uncorrected) brightnesses from Murthy, Henry, and 
Sujatha (2010).

At first glance, Fig. 7 looks very different from Fig. 5, but that is simply because overcorrected points (red spots; 
black spots) are clearly
present in Fig. 7, resulting in a lowest brightness that is clearly much too low.  I think it is most conservative to work 
with the uncorrected data, while keeping in mind that it contains contamination that could be significant
 at the lowest observed brightnesses.

In any case, comparison of Fig.~7 with Fig.~5 shows extremely good qualitative agreement, indicating that robust conclusions
are possible.

\section{Voyager observations}
Murthy and Henry (1999) reported the diffuse UV background at 1100 \AA\ for 430 locations observed using the UV 
spectrometers aboard the two Voyager spacecraft.  Now Professor Murthy has obtained and is processing or 
reprocessing a total of 1347 observations, including 917 previously unreduced observations from much farther out in the solar system, 
where the lower effects of solar activity are expected to result in much reduced contamination for these observations.  This should
allow us to either confirm or revise our earlier conclusions regarding the extremely low backgrounds that we have
previously reported.

\section{Conclusions}

With the availability of GALEX measurements of the spatial distribution of diffuse UV over the sky, we have entered a 
new era.  The observed spatial distribution
is incompatible with an origin of the diffuse UV background exclusively in dust-scattered starlight---a second 
component is required, at both high and low galactic latitudes.  The new component
cannot be cosmological (Henry 2010); it perhaps originates in weak interaction of the dark matter with the interstellar medium.

\begin{acknowledgements}
I am grateful to NASA for support, and to Franco Giovannelli for creating 
and leading the Frascati workshops. I thank Professor Jayant Murthy for helpful comments.\end{acknowledgements}

\bibliographystyle{aa}

\end{document}